\documentstyle[preprint,aps]{revtex}
\draft
\begin{document}
\title{Supercurrent noise in quantum point contacts}
\author{D. Averin and H. Imam}
\address{Department of Physics, SUNY at Stony Brook, Stony Brook,
NY 11794}
\maketitle

\begin{abstract}
Spectral density of current fluctuations in a short ballistic
superconducting quantum point contact is calculated for arbitrary
bias voltages $V$. Contrary to a common opinion that the supercurrent
flow in Josephson junctions is coherent process with no fluctuations,
we find extremely large current noise that is {\em caused} by the
supercurrent coherence. An unusual feature of the noise, besides its
magnitude, is its voltage dependence: the noise decreases with
increasing $V$, despite the fact that the dc current grows steadily
with $V$. At finite voltages the noise can be qualitatively understood
as the shot noise of the large charge quanta of magnitude $2\Delta
/V$ equal to the charge transferred during one period of Josephson
oscillations.

\end{abstract}

\vspace{3ex}

\narrowtext

It is well established that the supercurrent in classical
Josephson tunnel junctions does not fluctuate -- see, e.g.,
\cite{b77}. This result is consistent with the notion that the
supercurrent is the coherent dissipationless flow of Cooper pairs,
and therefore is quite appealing intuitively and has acquired the
status of a widespread opinion. There seems to be no reason
to suspect that the situation should be any different in mesoscopic
contacts with a few transverse modes, since, for instance, the noise
properties of contacts in the normal state are insensitive to the
number of transverse modes. Both in classical \cite{b1,b2} and
quantum \cite{b3,b4} normal point contacts the shot noise was
predicted to be suppressed due to the Pauli-principle correlations
between electrons, and such a suppression has been found in
experiments \cite{b5,b6}. This picture is modified only slightly in
point contacts between normal metals and superconductors, or in fully
superconducting point contacts at large voltages, where Andreev
scattering leads to partial reflection of electrons in the contact
region giving rise to a finite level of shot noise \cite{b2,b7}.
The noise in this case is associated only with the excess current,
and therefore represents only a small fraction of the full classical
shot noise.

The aim of this work is to suggest and prove by microscopic
calculation that contrary to intuitive expectation, the
supercurrent flow in quantum point contacts should generate
noise which is extremely large on the scale of the classical shot
noise. The noise arises from the interplay between quasiparticle
scattering and the supercurrent coherence. The latter amplifies
the randomness of the quasiparticle scattering by ``attaching'' to
each quasiparticle a large charge which is coherently transferred
through the point contact.

We begin by considering a single-mode ballistic contact between
two identical superconductors. Such contacts can be formed, for
instance, in superconductor/semiconductor heterosctructures and
exhibit quantization of the supercurrent \cite{b8}. The length $d$
of the contact is assumed to be much smaller than the coherence
length $\xi$, as well as the elastic and inelastic scattering lengths
in the superconductors, so that all scattering inside the contact
region can be neglected. We are interested in calculating the
spectral density of current $I(t)$ in the point contact:
\[ I(t)=\frac{ie\hbar}{2m}\sum_{\sigma}(\frac{\partial}{\partial z}-
\frac{\partial}{\partial z'})\psi^{\dagger}_{\sigma}(z,t)
\psi_{\sigma} (z',t')\mid_{z'\rightarrow z=0, t'\rightarrow t} \, , \]
where $z$ is the coordinate in the direction of the current flow and
$z=0$ corresponds to the position of the point contact. Taking $t$
and $t'$ in this definition to be on opposite branches of the
Keldysh contour, we express the current correlation function
\begin{equation}
 K_I(t_1,t_2)\equiv \frac{1}{2} \langle I(t_1)I(t_2)+I(t_2)I(t_1)
\rangle  - \langle I(t_1)\rangle \langle I(t_2) \rangle
\label{3} \end{equation}
as a product of two nonequilibrium Green's functions. In the
quasiclassical approximation this expression reduces to the
following \cite{b2}:
\begin{equation}
K_I(t_1,t_2) = -\frac{e^2}{8} \sum_{\pm} \mbox{Tr} [
g_{\pm}^>(t_1,t_2)\sigma_z g_{\pm}^<(t_2,t_1)\sigma_z +
g_{\pm}^<(t_1,t_2)\sigma_z g_{\pm}^>(t_2,t_1)\sigma_z ] \, ,
\label{4} \end{equation}
where the Green's functions $g$ are 2$\times$2 matrices in the
electron-hole space, Tr is taken over this space, and
$\sum_{\pm}$ is the sum over the two directions of propagation
through the contact. Here and below, $\sigma$'s denote Pauli
matrices.

{}From this point on, the calculation proceeds differently for
vanishing and finite bias voltages. At vanishing voltage, the
contact is in equilibrium and all Green's functions and current
correlation function (\ref{4}) depend only on the time difference
$\tau= t_1-t_2$. In the frequency domain the equilibrium functions
$g$ are:
\begin{equation}
g_{\pm}^>(\epsilon)=(1-f(\epsilon))\rho_{\pm} (\epsilon) \, , \;\;\;
g_{\pm}^<(\epsilon)=-f(\epsilon)\rho_{\pm} (\epsilon) \, ,
\label{41} \end{equation}
where $\rho(\epsilon) \equiv g_R(\epsilon)-g_A(\epsilon)$ is the
matrix of the density of states, and $f(\epsilon)$ is the equilibrium
Fermi distribution. Fourier transforming the correlation
function (\ref{4}) and using eq.\ (\ref{41}) we get the following
expression for the spectral density of current fluctuations
\begin{equation}
S_I (\omega ) = \frac{1}{2\pi} \int d\tau e^{-i\omega \tau} K(\tau) =
\frac{e^2}{32\pi^2\hbar } \sum_{\pm \omega, \, \pm} \int d\epsilon
f(\epsilon) (1-f(\epsilon \pm \hbar \omega))  \mbox{Tr} [\rho_{\pm}
(\epsilon) \sigma_z \rho_{\pm} (\epsilon \pm \hbar \omega) \sigma_z
] \, . \label{42} \end{equation}

We shall see below that in order to get a meaningful result for
$S_I(\omega)$ we need to keep a finite energy relaxation rate in the
model. We find the density $\rho (\epsilon)$ which includes a finite
relaxation rate generalizing the Kulik-Omelyanchuk theory \cite{b22}.
We perform the calculation in real time. For a short constriction
the retarded Green's function can be written as $g_{jR}^{(0)}
+g_{jR}$, where $g_{jR}^{(0)}$ is the spatially-uniform part of the
Green's function of the $j$th electrode:
\[ g_{jR}^{(0)} = \frac{1}{\delta} \left(
\begin{array}{cc} \tilde{\epsilon} & \tilde{\Delta} e^{i\varphi_j}
\\-  \tilde{\Delta} e^{-i\varphi_j} & - \tilde{\epsilon}
\end{array} \right) \, , \;\;\;\;\;
\tilde{\epsilon} =\epsilon+i\gamma_1\,, \;\;\; \tilde{\Delta}=
\Delta +i\gamma_2 \, ,\]
and $g_{jR}$ is the non-uniform part which satisfies the equation
\cite{b23,b24}:
\begin{equation}
\pm i v_F \frac{\partial g_R }{\partial z}= \delta [g^{(0)}_R
,g_R]  \, .
\label{44} \end{equation}
Here $\gamma_{1,2}$ are inelastic scattering rates, $\delta
\equiv (\tilde{\epsilon}^2 -\tilde{\Delta}^2)^{1/2}$ and $v_F$ is
the Fermi velocity. The non-uniform part of the Green's functions
should also satisfy two boundary conditions: $g \rightarrow 0$ at
$z\rightarrow \pm \infty$ (inside the electrodes), and
$g_1+g_1^{(0)}= g_2+g_2^{(0)}$ at $z=0$.

{}From eq.\ (\ref{44}) we obtain that the solutions decaying at
infinity have the following matrix forms:
\[ g_{1\, R}^{(\pm)} =A_1^{(\pm)}\left( \begin{array}{cc}
1 & (ae^{i\varphi_1})^{\pm 1} \\ -(ae^{i\varphi_1})^{\mp 1} & -1
\end{array} \right)  \, , \;\;\;\;
g_{2\, R}^{(\pm)} =A_2^{(\pm)}\left( \begin{array}{cc}
1 & (ae^{i\varphi_2})^{\mp 1} \\  -(ae^{i\varphi_1})^{\pm 1} & -1
\end{array} \right) \, , \]
where $a$ is the amplitude of Andreev reflection from the
superconductor:
\[ a(\epsilon) = (\tilde{\epsilon} -\delta)/\tilde{\Delta} \, . \]

Matching solutions in the two electrodes with the boundary condition
at $z=0$ we get the total Green's function at this point:
\begin{equation}
g_{R,\, \pm} = \frac{1}{\delta \cos (\varphi /2) \mp i
\tilde{\epsilon} \sin (\varphi /2) } \left( \begin{array}{c}
\tilde{\epsilon} \cos (\varphi /2) \mp i\delta \sin
(\varphi /2) \, , \;\;\; \tilde{\Delta}e^{i\varphi_{\Sigma}} \\
-\tilde{\Delta} e^{-i\varphi_{\Sigma}}\, , \;\;\; -\tilde{\epsilon}
\cos (\varphi /2) \pm i\delta \sin (\varphi /2) \end{array} \right)
\, .
\label{45} \end{equation}
where $\varphi_{\Sigma} =(\varphi_1+\varphi_2)/2$. The advanced
function $g_A$ can be expressed in
terms of $g_R$, $g_A(\varphi_{\Sigma}) =
- g_R^*(-\varphi_{\Sigma})$. Using this relation and eq.\ (\ref{45}),
we see that although the Green's functions are not gauge-invariant
(i.e., depend on $\varphi_{\Sigma}$), the current spectral density
(\ref{42}) is gauge invariant. Therefore, to simplify the notations,
we take below $\varphi_{\Sigma}=0$ without loss of generality.

In the most interesting limit of small energy relaxation,
$\gamma_{1,2} \ll \Delta/\hbar$, there are two distinct regions of
energy where the density of states $\rho$ is non-vanishing: one
is $\mid \epsilon \mid > \Delta$, and the other is the vicinity of
the two subgap states with energies $\pm \epsilon_0 = \pm \Delta
\cos (\varphi/2)$. Expanding eq.\ (\ref{45}) in small deviations
from $\pm \epsilon_0$, we get that in the subgap region the density
of states is:
\begin{equation}
\rho_{\pm} (\epsilon) = 2 \mbox{Re} [g_{R,\, \pm} (\epsilon)] =
\frac{\gamma (\epsilon_0) \Delta \sin (\varphi/2)}{ (\epsilon \mp
\epsilon_0)^2 +\gamma^2 (\epsilon_0)/4 }\left( \begin{array}{cc}
1 & \pm 1 \\ \mp 1 & -1 \end{array} \right) \, .
\label{46} \end{equation}
Using this relation in eq.\ (\ref{42}), we get the subgap
contribution to the current noise at low frequencies $\omega
\sim \gamma$:
\begin{equation}
S^{(1)}_I(\omega) = \left( \frac{I_0 \sin (\varphi/2)}{\cosh
(\epsilon_0/2T)} \right) ^2 \frac{\gamma (\epsilon_0) }{2\pi
(\omega^2+ \gamma^2 (\epsilon_0)) } \, ,
\label{47} \end{equation}
where $I_0\equiv e\Delta/\hbar$, and
\begin{equation}
\gamma(\epsilon)= 2(\gamma_1(\epsilon)-\frac{\epsilon}{\Delta}
\gamma_2 (\epsilon)) = \alpha \int d\epsilon' \frac{\Theta
(\epsilon'^2 -\Delta^2)}{\sqrt{\epsilon'^2 -\Delta^2}}
\frac{(\epsilon-\epsilon')^3 \cosh(\epsilon/2T)}{
\sinh ((\epsilon-\epsilon')/2T) \cosh(\epsilon'/2T)} \, .
\label{48}  \end{equation}
Here $\alpha$ is a constant determined by the parameters of
electron-phonon interaction.

Equation (\ref{47}) shows that there is a very large low-frequency
noise associated with the supercurrent flow through a quantum
point contact. Although the total noise intensity decreases with
decreasing temperature, its zero-frequency density can actually
increase at low temperatures because of the rapid decrease of
$\gamma$. The noise has a very simple time-domain interpretation.
Equation (\ref{47}) implies that the current which is considered
to be a ``dc'' supercurrent does not flow continuously; rather it
is a stochastic process with a typical realization shown in the
inset in Fig.\ 1. Because of the quasiparticle exchange between
the bulk electrodes and the two subgap states $\pm \epsilon_0$
localized in the point contact, the system jumps between the
state carrying positive current $I_0 \sin (\varphi /2)$ and the
state carrying negative current $-I_0 \sin (\varphi /2)$ with an
average rate $\gamma(\epsilon_0)$. The jumps occur in such a way that
the probabilities of finding a positive and negative current are,
respectively, $f(-\epsilon_0)$ and $f(\epsilon_0)$. It is
straightforward to check that the spectral density of such a process
is indeed given by eq.\ (\ref{47}). Thus, the noise (\ref{47}) is
the two-level noise which is an inherent part of the supercurrent
thermalization in the quantum point contacts.

Equations (\ref{42}) and (\ref{45}) also determine the noise due
to the the states above the gap. As we will see below, this
part of the noise has a much less singular behavior in $\gamma$
than the subgap noise (\ref{47}). Therefore in order to calculate it
we can limit ourselves to $\gamma=0$. In this case we obtain from
eqs.\ (\ref{42}) and (\ref{45}):
\[ S_I^{(2)}(\omega) = \frac{e^2}{4\pi^2 \hbar}\int d \epsilon [1-
\tanh (\frac{\epsilon}{2T})\tanh (\frac{\epsilon+\hbar \omega}{2T} )]
u(\epsilon)u(\epsilon+\hbar \omega) ( \mid \epsilon \mid
\mid \epsilon+  \hbar \omega \mid +  \]

\vspace*{-3ex}

\begin{equation}
\mbox{sgn} (\epsilon) \mbox{sgn} (\epsilon+\hbar \omega)
\Delta^2 \cos^2 (\varphi/2))  \, , \;\;\;\; u(\epsilon)
\equiv \frac{(\epsilon^2-\Delta^2)^{1/2} \Theta (\epsilon^2-\Delta^2
)}{ \epsilon^2 - \Delta^2 \cos^2 (\varphi/2)} \, .
\label{49} \end{equation}
An interesting feature of the noise (\ref{49}) is that it is
phase-dependent, despite the fact that the states above the gap do
not contribute to the average supercurrent. For $T\gg \Delta$ and
$\omega=0$, eq.\ (\ref{49}) gives:
\begin{equation}
S_I^{(2)}(0) = \frac{e^2}{\pi^2 \hbar } [T - \frac{ \Delta}{2}
(\cos^3 \frac{\varphi}{2} \ln \left| \frac{1+\cos (\varphi/2)}{
1-\cos (\varphi/2)} \right| -1 )] \, .
\label{50} \end{equation}
The first term in this expression is the regular Nyquist noise as
in the normal state, while the second term is the phase-dependent
correction associated with the gap. Because of the usual BCS
singularity in the density of states, the second term diverges weakly
as $\varphi \rightarrow 0$. This divergence is removed by $\gamma$;
at small but finite $\gamma$, the logarithm is limited by
$\ln(\Delta/\gamma)$. At $\varphi =0$, $S_I^{(2)}(\omega)$ has
a similar weak singularity as a function of $\omega$,
\[ S_I^{(2)}(\omega) = \frac{e^2}{\pi^2 \hbar } [T -\frac{\Delta}{2}
\ln \frac{\Delta}{\omega}] \, .\]

The last component of noise comes from the ``interference'' of
the subgap states and states above the gap in eq.\ (\ref{42}).
In the limit $\gamma \rightarrow 0$ we get from this equation and
eqs.\ (\ref{45}) and (\ref{46}):
\begin{equation}
S_I^{(3)}(\omega) = \frac{e^2\Delta }{4\pi\hbar^2 \omega}\sum_{\pm}
\Theta ((\epsilon_0\pm \hbar \omega)^2-\Delta^2) ((\epsilon_0\pm
\hbar \omega)^2-\Delta^2)^{1/2} [1- \tanh (\frac{\epsilon_0}{2T})
\tanh (\frac{\epsilon_0 \pm \hbar \omega}{2T})] \, .
\label{51} \end{equation}
Comparison of eqs.\ (\ref{50}) and (\ref{51}) with eq.\ (\ref{47})
shows that the subgap noise dominates at low frequencies as long as
the temperature is not too large on the scale of $\Delta$, i.e.,
$T\ll \Delta^2 /\gamma$.

Total spectral density of current fluctuation $S_I(\omega)$ as a
function of frequency calculated numerically from eqs.\ (\ref{42}) and
(\ref{45}) is shown in Fig.\ 1 for several values of the Josephson
phase difference $\varphi \in [0,\, \pi]$. The $S_I(\omega)$ for
$\varphi \in [\pi,\, 2\pi]$ can be found from the relation
$S_I(\omega, \varphi)= S_I(\omega, 2\pi-\varphi)$ which follows
from (\ref{42}) and (\ref{45}).  Figure 1 shows that the subgap noise
(\ref{47}) indeed dominates the spectrum even at not too small
$\gamma$. We also see that due to the interference term (\ref{51})
the high-frequency threshold of $S_I(\omega)$ at low temperatures
shifts down from $2\Delta/\hbar$ with increasing $\varphi$. In
particular, at $\varphi =\pi$, the noise at large frequencies starts
at $\Delta/\hbar$ in accordance with eq.\ (\ref{51}).

Now we turn to finite voltages. In this case, all observable
quantities like the current
correlation function (\ref{4}) are periodic in time $t=(t_1+t_2)/2$
with the period of the Josephson oscillations $T_0= \pi \hbar /eV$.
This implies that the Green's functions $g$ should be periodic with
period $2T_0$, since their off-diagonal elements can change sign
when shifted by $T_0$. Therefore, $g$ can be expanded as a Fourier
series:
\[ g(t_1,t_2)= \sum_k g_k(\tau) e^{-ikeVt/\hbar} \, , \]
where $\tau\equiv t_1-t_2$. Averaging the correlator (\ref{4}) over
$t$ and Fourier transforming it with respect to the time difference
$\tau$ as in the static case, we get the following expression for
the spectral density of current fluctuations:
\begin{equation}
S_I (\omega ) = -\frac{e^2}{16 \pi^2 \hbar} \sum_{k,\, \pm \omega}
\int d\epsilon \mbox{Tr} [ g_k^>(\epsilon ) \sigma_z g_{-k}^<(\epsilon
\pm \hbar \omega ) \sigma_z ] \, .
\label{5} \end{equation}
In eq.\ (\ref{5}) we have taken into account the fact that summation
over the two
directions of propagation in the correlator (\ref{4}) is equivalent,
for a symmetric contact, to summation over positive and negative bias
voltages $\pm V$. Since the noise is an even function of $V$, this
summations only contributes a factor of 2.

{}From the known solution for Green's functions of a short
superconducting point contact \cite{b14,b10} and standard relations
between the different Green's functions:
\[g^> = \frac{1}{2}(g^K+g^R-g^A) \, , \;\;\; g^< = \frac{1}{2}
(g^K -g^R+g^A) \, , \]
we obtain the Fourier components $g_k$:
\begin{equation}
g_k=p_k \sigma_z +q_k i\sigma_y \, .
\label{6} \end{equation}
Here $p_k$ is non-vanishing for even $k$ and:
\begin{equation}
p_k^>(\epsilon) = 2(1-F(\epsilon -eV/2)) A(\epsilon) \, ,\;\;\;
p_k^<(\epsilon) = -2F(\epsilon -eV/2)A(\epsilon) \, ,
\;\;\; k\geq 0 \, , \label{7} \end{equation}
\[ A(\epsilon) \equiv \prod_{l=1}^k a(\epsilon+leV-eV/2) \, , \]
while $q_k$ is non-vanishing for odd $k$ and is given by the same
eq.\ (\ref{7}) with odd $k$. For negative $k$, $p_k=p_{-k}^*$,
and $q_k=q_{-k}^*$. Function $F$ has the meaning of non-equilibrium
distribution of quasiparticles in the point contact:
\begin{equation}
F(\epsilon)= f(\epsilon)+ \sum_{n=0}^{\infty} \prod_{m=0}^n \mid
a(\epsilon-meV) \mid^2 [f(\epsilon-(n+1)eV) - f(\epsilon-neV)] \, ,
\label{8} \end{equation}
Combining eqs.\ (\ref{6}) and (\ref{7}) with eq.\ (\ref{5}) we get
a final expression for $S_I (\omega )$:
\begin{equation}
S_I (\omega ) = \frac{e^2}{2 \pi^2 \hbar} \sum_{ \pm \omega}
\int d\epsilon F(\epsilon)(1-F(\epsilon \pm \hbar \omega))
[1+2\mbox{Re} \sum_{k=1}^{\infty} \prod_{l=1}^k a(\epsilon+leV)
a^*(\epsilon+leV\pm \hbar \omega) ]\, .
\label{10} \end{equation}

Equation (\ref{10}) together with eqs.\ (\ref{42}) and (\ref{45})
are the main technical results of our work. Combined, these equations
give the spectral density of current fluctuations in a short ballistic
constriction between two identical superconductors at arbitrary
voltages. As we can expect from our calculations for $V=0$, the
most interesting limit at finite voltages is $V\ll \Delta/e$. In this
case eq.\ (\ref{10}) can be simplified further. Expanding the
amplitudes $a(\epsilon)$ of Andreev reflection in small relaxation
rates $\gamma_{1,2}$ and replacing the sums with the integrals in eqs.\
(\ref{8}) and (\ref{10}) we obtain the spectral density of current
fluctuations at low frequencies, $\omega \ll \Delta/\hbar$:
\[ S_I (\omega ) =  \frac{e}{\pi^2 \hbar V} \sum_{\pm \omega}
\int_{- \Delta}^{\Delta} d\epsilon F(\epsilon)(1- F(\epsilon \pm\hbar
\omega)) \int^{\Delta}_{\epsilon }d\epsilon' \exp \{ -
\int_{\epsilon }^{\epsilon'}\frac{d\nu \hbar \gamma(\nu)}{eV
\sqrt{\Delta^2-\nu^2}} \} \times \]

\vspace*{-3ex}

\begin{equation}
\cos ( \frac{\hbar \omega}{eV} [\arccos (\frac{\epsilon
+\epsilon'}{\Delta}) - \arccos \frac{\epsilon}{\Delta} ] ) \, ,
\label{11} \end{equation}
where the quasiparticle distribution function reduces to
\begin{equation}
F(\epsilon) = f(\epsilon ) - \int_{-\Delta}^{\epsilon }d\epsilon'
\frac{\partial f}{\partial \epsilon'}\exp \{ -\int_{ \epsilon' }^{
\epsilon }\frac{ d \nu \hbar \gamma (\nu)}{eV \sqrt{\Delta^2-\nu^2}}
\} \, .  \label{111} \end{equation}

Equation (\ref{11}) is still a very general result which gives
the spectral density of current fluctuations for arbitrary relation
between the bias voltage, temperature, frequency, and energy
relaxation rate. At large temperatures $T\gg \Delta$ the relaxation
rate $\gamma$ (\ref{48}) and distribution function $F$ are constant
for energies inside the gap, and the zero-frequency spectral density
(\ref{11}) is:
\begin{equation}
S_I(0)= \frac{e\Delta^2}{4\pi^2 \hbar V} \, \frac{\pi \lambda
(1+\lambda^2) +2+2e^{-\pi \lambda} }{(1+\lambda^2)^2} \, ,
\;\;\;\; \lambda \equiv \frac{\hbar \gamma}{eV} \, .
\label{12} \end{equation}
The noise (\ref{12}) has an unusual voltage dependence; it decreases
monotonically with increasing voltage despite the fact that, as can
be shown, the average current grows with voltage. Although the
monotonic decrease of the noise intensity with the voltage is a
characteristic feature of large temperatures $T\gg \hbar \gamma$,
noise always decreases with voltage at $V\gg \hbar \gamma/e$ (see inset
in Fig.\ 2). Indeed, in this case, eq.\ (\ref{111}) shows that
$F(\epsilon)= F(\Delta)$, and we get for the spectral density
(\ref{11}):
\begin{equation}
S_I(\omega) = \frac{e\Delta^2 }{2\pi^2 \hbar \cosh^2 (\Delta/2T) V}\,
\frac{1+ \cos (\pi \hbar \omega/eV)}{(1-(\hbar \omega /eV )^2)^2}\, .
\label{13} \end{equation}
At small voltages  $V\ll \hbar \gamma/e$, and temperatures $T\ll
\Delta$, the noise is independent of the voltage $V$:
\begin{equation}
S_I(0) = \frac{2 e^2\Delta T}{\pi^2 \hbar^2 \gamma(0) }\, .
\label{14} \end{equation}
Equations (\ref{13}) and (\ref{14}) describe two sides of the noise
peak with the maximum at $V\simeq \hbar \gamma/e$. The exact shape
of this peak depends on the energy dependence of the relaxation
rate $\gamma$ and is shown in the inset in Fig.\ 2 for several
temperatures in the approximation of energy-independent $\gamma$.
The curves were calculated numerically from eq.\ (\ref{10}).
The main part of Fig.\ 2 shows the zero-frequency spectral density
of current fluctuations at arbitrary voltages. It illustrates a
transition from the noise peak at small voltages to the large
voltage regime where $S_I(0)$ saturates at $4e^2\Delta/15 \pi^2
\hbar$.

The fact that the noise at small voltages is very large on the
scale of the regular shot noise, together with an unusual voltage
dependence, reflects an unusual physical mechanism of the noise.
For $V\gg \hbar \gamma/e$ this mechanism can be described
qualitatively as follows. The giant noise arises since each
quasiparticle getting into the constriction region with energy
equal to one of the gap edges generates an avalanche of Andreev
reflections before it can escape out of the constriction by climbing
up or down in energy to the opposite edge of the energy gap. The
number of generated Andreev reflections is $2\Delta/eV$, so that
each quasiparticle causes a coherent transfer through the
constriction of a charge quantum of magnitude $2\Delta /V$. For small
voltages $V\ll \Delta/e$ this is much larger than the charge of
individual Cooper pairs. In this way, the randomness of the
quasiparticle scattering (quasiparticles get inside the energy gap
with probability $f(-\Delta)$ from one electrode and with
probability $f(\Delta)$ from the opposite electrode) is amplified.
Therefore, the noise described by eq.\ (\ref{13}) can be interpreted
as the shot noise of these large charge quanta, and is in fact the
{\em consequence} of the coherence of the supercurrent flow through
the point contact.

It is interesting to note that this picture in the energy domain has
a ``dual'' formulation in the time domain in terms of the
non-equilibrium  occupation of the two subgap states which are
responsible for the dc supercurrent. In particular, the avalanche
of $2\Delta/eV$ Andreev reflections triggered by a quasiparticle
corresponds in the time domain to one period of Josephson oscillation,
during which the supercurrent $I_0 \sin (\varphi/2)$ carries the
charge $2\Delta /V$ through the point contact. More generally, all
small-voltage results for the spectral density of current
fluctuations obtained above (eqs.\ (\ref{12}) -- (\ref{14}) for the
ac regime and eq.\ (\ref{47}) for the dc regime) can also be
obtained from a purely classical rate equation \cite{b10} for the
non-equilibrium occupation probabilities of the two subgap
states. An advantage of such simple classical approach is the
possibility to generalize it straightforwardly to time-dependent
voltages and arbitrary bias conditions of the point contact. Of
course an advantage of the fully microscopic approach used in this
work is that it gives spectral density of current fluctuations for
frequencies and bias voltages that are arbitrary on the scale of
$\Delta$.

The simple time-domain interpretation of the supercurrent noise
at small bias voltages allows us to propose a simple generalization
of our results for ballistic junctions to junctions with arbitrary
transmission coefficient $D$. Indeed, making a very natural
assumption that we have the same exchange of quasiparticles between
the two subgap states localized in the point contact and the bulk
superconductors, we get immediately for $V=0$ (cf.\ eq.\ (\ref{47})):
\begin{equation}
S_I(\omega) = \frac{1}{2\pi} \sum_{k=1}^N \left( \frac{I_k
(\varphi)}{\cosh (\epsilon_k/2T)} \right) ^2 \frac{\gamma
(\epsilon_k) }{ \omega^2+ \gamma^2 (\epsilon_k) } \, ,
\label{15} \end{equation}
Here $N$ is the number of propagating transverse modes in the
junction, $I_k(\varphi )\equiv (e\Delta/2\hbar) D_k \sin(\varphi)/
[1-D_k\sin^2(\varphi/2)]^{1/2}$ coincides at $T=0$ with the
supercurrent carried by the $k$th mode, and $\epsilon_k =
\Delta [1-D_k\sin^2(\varphi/2)]^{1/2}$ \cite{b11,b12}. Since
fluctuations of the current in all modes are independent, the
current noise $S_I(\omega)$, expressed in terms of the total
supercurrent, decreases roughly as $1/N$ with increasing number
$N$ of transverse modes. Therefore the giant small-voltage noise
discussed in our work disappears in the classical limit $N
\rightarrow \infty$. However, it can be very important in quantum
point contacts with few transverse modes, and even in regular
tunnel Josephson junctions of ultrasmall area.

In conclusion, we have shown that the supercurrent flow in quantum
point contacts leads to a giant current noise both in the regime of
dc and ac Josephson effects. The noise arises from the interplay of
quasiparticle scattering and coherence of the supercurrent flow and
has an unusual temperature and voltage dependence.

The authors gratefully acknowledge a suggestion of K. Likharev
that stimulated this work, and discussions with A. Korotkov
and A. Zaitsev. This work was supported in part by DOD URI through
AFOSR grant \# F49620-92-J0508.

\figure{Figure 1.} Spectral density of current fluctuations
in a short single-mode constriction between two superconductors
in the regime of the dc Josephson effect. From bottom to top,
the curves correspond to $\varphi =0; \, \pi/2; \,\pi$. The peak at
low frequencies is the subgap contribution \protect (\ref{47}).
The inset shows a typical realization of the ``dc'' supercurrent
as a function of time at non-vanishing temperatures. The sign of
the current switches randomly with the characteristic rate
$\gamma$ giving rise to the low-frequency peak in spectral
density.

\figure{Figure 2.} Zero-frequency density of current fluctuations
in the ac regime as a function of the bias voltage at zero
temperature. The inset shows a blow-up of the small-voltage peak
at (from bottom to top) $T=0;\, 0.1\Delta; \, 0.2\Delta$.

\end{document}